\documentclass[aps,prl,twocolumn,superscriptaddress]{revtex4}
\usepackage{graphicx}
\usepackage{mathrsfs}
\usepackage{bm}
\usepackage{amsmath}
\usepackage{dcolumn}
\usepackage{epstopdf}
\usepackage{epsfig}
\usepackage{dsfont}
\usepackage{amssymb}
\usepackage{tabularx}
\usepackage{array}
\usepackage{float}
\usepackage{colordvi}
\usepackage{color}

\begin{document}
\title{Quantum Phases of Kagome Electron System with Half-Filled Flat Bands}
\author{Yafei Ren}
\affiliation{ICQD, Hefei National Laboratory for Physical Sciences at Microscale, CAS Key Laboratory of Strongly-Coupled Quantum Matter Physics, and Department of Physics, University of Science and Technology of China, Hefei, Anhui 230026, China}
\affiliation{Department of Physics and Astronomy, California State University, Northridge, California 91330, USA}
\author{Hong-Chen Jiang}
\affiliation{Stanford Institute for Materials and Energy Sciences, SLAC National Accelerator Laboratory and Stanford University, Menlo Park, California 94025, USA}
\author{Zhenhua Qiao}
\affiliation{ICQD, Hefei National Laboratory for Physical Sciences at Microscale, CAS Key Laboratory of Strongly-Coupled Quantum Matter Physics, and Department of Physics, University of Science and Technology of China, Hefei, Anhui 230026, China}
\author{D. N. Sheng}
\affiliation{Department of Physics and Astronomy, California State University, Northridge, California 91330, USA}
	
\begin{abstract}
  We study the quantum phase diagram of electrons on kagome lattice with half-filled lowest flat bands. To understand the interplay between repulsive interactions, magnetism, and band topology, we adopt an extended $t$-$J$ model, where the hopping energy $t$, antiferromagnetic Heisenberg interaction $J$, and short-range neighboring Hubbard interaction $V$ are considered. In the weak $J$ regime, we identify a fully spin-polarized phase, which can further support the spontaneous Chern insulating phase driven by the short-range repulsive interaction $V$. We find that this phase is independent of the spin orientation in contrast to the spin-orbit coupling induced Chern insulator that disappears with in-plane ferromagnetism constrained by symmetry. 
  Such symmetry difference provides a criterion to distinguish the physical origin of topological responses in kagome metals. 
  As $J$ increases gradually, we find that the ferromagnetic topological phase is suppressed, which first becomes partially-polarized and then enters a non-magnetic phase with nematic charge density wave.
  In the end, we discuss the potential experimental observations of our theoretical findings.
\end{abstract}
\date{\today}

\maketitle

\textit{Introduction---.} The geometrical frustration of kagome lattice has attracted much attention in the past few decades~\cite{KAFM_PD_15, Rev_QSL_17, RMP_QSL_17, Rev_QSL_20, Comment_Kagome_20, Kagome_AF_SpinLiquid_Jiang_08, Kagome_XXY_15, Kagome_NNNsoc_TI_09, 6thFilled_Hole_JianxinLi_14, 6thFilled_Hole_Thomale_13, Kagome_Hubbard_Soc_15, Kagome_Hubbard_Half_14, Kagome_Hubbard_Half_bondformation_16, Kagome_DopedHalfFilling_tJ_13, Kagome_DopedHalfFilling_Wang_13, Kagome_Spin_DipoleInt_17, Material_QSL_20,Kagome_FlatHalf_Pollmann_08}. At half filling, the frustration of anti-ferromagnetic spins can lead to quantum spin liquid phases and other exotic magnetic orders~\cite{KAFM_PD_15, Rev_QSL_17, RMP_QSL_17, Rev_QSL_20}.
Besides, the frustration of electronic hopping causes a flat band that shows quadratic crossing with Dirac bands~\cite{HopFrustration_08}. For a semi-metal where the Fermi energy lies at the Dirac or quadratic crossing points, the presence of strong electronic interaction can open a band gap at the fermi points hosting nontrivial topology~\cite{QAHE_QBC_Sun_09, QAHE_QBC_Sun_ColdAtom_12, QAHE_KagomeDecGra_MF_10, QAHE_Kagome_NN_ColdAtom_18, QAHE_Lieb_Tsai_15, QAHE_QBC_Kagome_MF_10, Kagome_2thirdFilling_Pollmann_14, QAHE_16_PRL_Zhu, QAHE_Gong_18, QAHE_QBC_Zeng18, QAHE_Kagome_18,QAHE_Lukas_20, Kagome_DiracFilling_14}.
Recently, metallic kagome lattices of transition metal systems are discovered experimentally, which provide a novel platform to investigate the intriguing physics from the interplay between strong repulsive interaction, magnetism, and topology~\cite{18_Nature_Ye_Dirac, 18_Ye_dHvA_Fe3Sn2, 18_Nature_Hasan_Fe3Sn2, 19_Kang_FeSn_DiracFlat, 19_Kang_FeSn_ElecMagneticProp, 18_NatPhys_GiantAHE_Co3Sn2S2, 19_NatPhys_FlatBandMag_Co3Sn2S2, 18_NatComm_AHE_Co3Sn2S2, 19_APL_AHE_InPlane_Co3Sn2S2, 19_YCr6Ge6_FlatBand, 20_CoSn_FlatBand_NonMag, Fe3Sn2_BfieldTuning_Hasan_19, CoSnS_InPlaneAFM_Hasan_19, CoSnS_ExchangeBias_NC_19}.

The kagome lattices formed by transition metal atoms exhibit characteristic single-particle band structures, including quasi-two-dimensional Dirac bands~\cite{18_Nature_Ye_Dirac, 18_Ye_dHvA_Fe3Sn2, 18_Nature_Hasan_Fe3Sn2, 19_Kang_FeSn_DiracFlat} and flat bands near the Fermi energy~\cite{19_NatPhys_FlatBandMag_Co3Sn2S2, 19_Kang_FeSn_DiracFlat, 19_Kang_FeSn_ElecMagneticProp, 19_YCr6Ge6_FlatBand, 20_CoSn_FlatBand_NonMag}. Nontrivial band topology is observed by measuring anomalous Hall effect and orbital magnetic moments~\cite{18_Nature_Ye_Dirac, 18_Ye_dHvA_Fe3Sn2, 18_Nature_Hasan_Fe3Sn2, CoSnS_ExchangeBias_NC_19, 19_Kang_FeSn_DiracFlat, 19_NatPhys_FlatBandMag_Co3Sn2S2, 18_NatPhys_GiantAHE_Co3Sn2S2, 18_NatComm_AHE_Co3Sn2S2, 19_APL_AHE_InPlane_Co3Sn2S2}. 
These topological responses are closely related to the magnetic phases and strong electronic interactions as indicated by the experiments~\cite{18_Nature_Ye_Dirac, 18_Ye_dHvA_Fe3Sn2, 18_Nature_Hasan_Fe3Sn2, CoSnS_ExchangeBias_NC_19, 19_NatPhys_FlatBandMag_Co3Sn2S2}. 
However, the role of interaction, and its interplay with magnetism and topological band structure remain not clearly understood
in kagome metal systems.

\begin{figure}
	\includegraphics[width=8.5 cm]{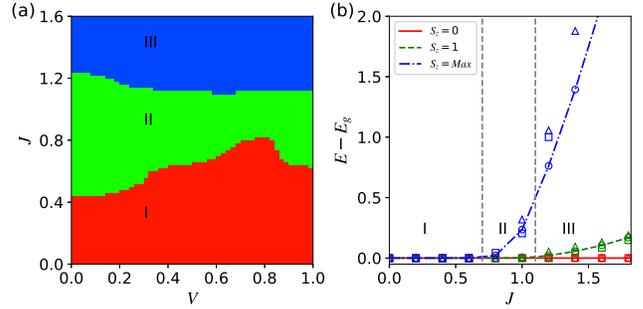}
	\caption{(a) Phase diagram vs on-site repulsion-interaction strength $V$ and exchange-interaction strength $J$. Phases I, II, and III are fully spin polarized spontaneous Chern insulator phase, partially polarized intermediate phase, and non-polarized phase with a rotation-symmetry breaking charge density wave, respectively. (b) Lowest energy level at each sector of azimuthal spin polarization $S_z$ vs exchange-interaction strength $J$ at $V=0.6$. Circle, triangle, and square stand for the system with $N_s=36$, $48$ and $72$, separately. The phase diagram is calculated for $N_s=3\times 3 \times 4=36$ site system. The first and second nearest neighbor interactions are $V_1=2V_2=V$. The dashed lines indicate the phase boundaries. }
	\label{PhaseDiagram}
\end{figure}

In this Letter, we theoretically investigate the interplay of topological phases and magnetism in the presence of strong correlation based on the extended $t$-$J$ model on kagome lattice, where $J$ is the antiferromagnetic exchange interaction between the nearest neighbor sites and $V$ is the repulsive interaction up to the second nearest neighbor sites. For electrons at the half-filling of the lowest flat band, we identify three phases including a fully spin-polarized phase  with a spontaneous quantized Chern number, a partially spin-polarized intermediate phase, and a spin-unpolarized phase with symmetry breaking nematic charge density wave as the exchange interaction $J$ increases. The spontaneous Chern insulator emerges in the fully spin-polarized phase at a small $J$ and is independent of the direction of spin polarization, which distinguishes itself from the traditional spin-orbit coupling (SOC) induced Chern insulator. Along with the increase of $J$, the system first undergoes a quantum phase transition to the partially spin-polarized phase and finally to a non-magnetic phase that exhibits a charge-density-wave order. These different quantum phases exhibit distinct responses by applying an external Zeeman field.

\textit{Model and methods---.} We consider a kagome lattice  of spinful fermions described by an extended $t$-$J$ model with the Hamiltonian written as:
\begin{align}
\label{Ham}
H &= t\sum_{\langle ij \rangle, \alpha}c_{i,\alpha}^\dagger c_{j,\alpha} + J\sum_{\langle ij  \rangle} ({\bm S}_i \cdot {\bm S}_j - \frac 1 4 n_i n_j )\nonumber \\
  &+ V_1\sum_{\langle ij \rangle}n_{i} n_{j} + V_2\sum_{\langle \langle ij \rangle \rangle}n_{i} n_{j}, 
\end{align}
where $c_{i,\alpha}^\dagger$ ($c_{i,\alpha}$) is the creation (annihilation) operator of a fermion with spin $\alpha=\{\uparrow, \downarrow \}$ at site $i$ and $n_i = \sum_\alpha c_{i,\alpha}^\dagger c_{i,\alpha}$ is the particle number operator.
$\bm{S}_i=(S_i^x, S_i^y, S_i^z) = \frac{1}{2} \sum_{\alpha,\beta}c_{i,\alpha}^\dagger \sigma_{\alpha,\beta}^{x,y,z} c_{i,\beta}$ is the spin operator with $\sigma^{x,y,z}$ being Pauli matrices.
The hopping term  $t$ is set to be the energy unit, which makes the lowest energy band flat and quadratically touching with the middle one.
The second term represents the exchange interaction with strength $J>0$ (antiferromagnetic type) between each pair of nearest neighbor sites.
The third and fourth terms are the repulsive interactions between electrons of first ($\langle\ldots\rangle$) and second ($\langle\langle\ldots\rangle\rangle$) nearest neighbors with strengths $V_1$ and $V_2$, respectively. The Hilbert space is constrained by the no-double occupancy condition, $n_i\leq 1$, which corresponds to the $U=\infty$ limit.

We focus on the one-sixth filling case in a finite system of $N_{x} \times N_{y}$ unit cells with total number of sites $N_{s} = 3\times N_{x} \times N_{y}$ and the number of fermions $N_{e} = 2N_{s}/6$. Without loss of generality, we take $V_1 = 2V_2 = V$ in our calculations. The Zeeman field $B_{\rm Z}$ and SOC $t_{\rm SO}$ are set to zero unless otherwise noted. To characterize the ground states of the system with interactions, we employ the finite density matrix renormalization group (DMRG) algorithm~\cite{DMRG_White_92, DMRG_Revisit_08, QAHE_15_PRB_Zhu, QAHE_14_SciRep_Gong} on cylinder geometry, where the boundary is open (periodic) along $x$ ($y$)-direction, respectively.
In DMRG calculations, we set $N_y$ up to 4 unit cells (8 lattice sites) and keep the DMRG states up to $M=12000$ to guarantee a good convergence (with the truncation error smaller than $10^{-5}$).

\textit{Phase diagram---.} As each spin component is conserved in our model, the ground states are calculated in sub-Hilbert space with total azimuthal spin $S_z$ ranging from $0$ to $S_{max}$ with $S_{max}=N_e/2$ (results are symmetric about positive and negative total $S_z$).
For a system of $N_s=3\times 3\times 4=36$, we numerically calculate the ground-state energies for different $S_z$ sector in the parameter space spanned by extended Hubbard interaction $V$ and exchange interaction $J$. We identify three phases as shown in Fig.~\ref{PhaseDiagram}(a) according to the polarization of the ground state.
In phase-I, the system is fully spin-polarized, i.e., the ground states of different spin $S_z$ ranging from $-S_{max}$ to $S_{max}$ are all degenerate with a total spin  $S=N_e/2$. Our results suggest that the spontaneous ferromagnetization in this system is very strong, which can survive to finite antiferromagnetic coupling $J\sim 0.4$.
Interestingly, an intermediate interaction strength $V\sim 0.8$ can further enlarge the regime of the fully polarized phase.
This result is complementary to the kinetic ferromagnetism in Ref.~\onlinecite{Kagome_FlatHalf_Pollmann_08} that is insensitive to the sign of hopping energy and focus on the regime of $|t|\ll V$.
Besides, the weak $J$ limit corresponds to a large $U$ and the ferromagnetic ground state agrees with the graph-theory analysis of $t$-$U$ model on kagome lattice with Hubbard $U>0$~\cite{Mielke_92}. As $J$ increases, however, we find that the ferromagnetic phase is suppressed.
For intermediate $J$, the ground state jumps from $S=N_e/2$ to a partially spin-polarized state with a smaller total $S$ driven by antiferromagnetic coupling, which is illustrated as phase-II. When $J$ further increases, the partially polarized phase also becomes excited states and the ground state lies in the spin sector of total $S=0$ labeled as phase-III in Fig.~\ref{PhaseDiagram}.

\begin{figure}
	\includegraphics[width=8.5 cm]{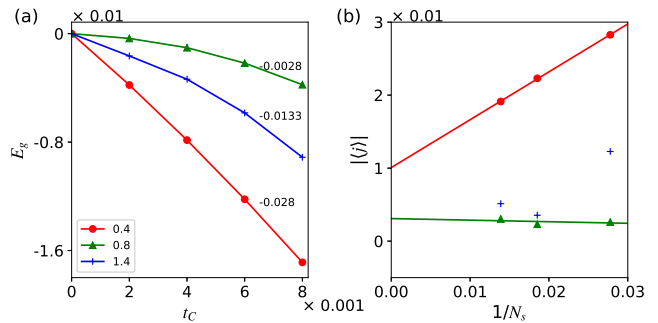}
	\caption{(a) Variation of ground state energy due to the presence of nonzero SOC $t_{\rm C}$ at different $J$ for $N_s=36$ sites. The slope of each curve at $t_{\rm C}=0$ reprents the average current $\langle j\rangle$ of the system as labelled by the number nearby. (b) Dependence of current $\langle j \rangle$ as function of exchange interaction strength $J$. Blue curve is the result for $N_s=36$. Green squares and red triangles are results for $N_s=54$ and $N_s=72$ sites. Inset: Scaling of current at $J=0.4$ for different sizes.}
	\label{Current}
\end{figure}

We further compare results from different system sizes to show the robustness of the quantum phase diagram.
As shown in Fig.~\ref{PhaseDiagram}(b) at fixed $V=0.6$, we present the energy difference $E_g^{Sz}-E_g^0$ between ground state energy at different $S_z=0,1,S_{max}$ as a function of $J$, for three different system sizes $N_s=3\times4\times 3=36$, $3\times4\times 4=48$ and $3\times 3\times 8=72$ sites, respectively. The same energy evolution with $S_z$ is identified for these different systems.
For a smaller $J$, the ground state has a total $S=S_{max}$ and the lowest energies from each $S_z$ sector
has the same energy, representing a $2S+1$ magnetic degeneracy. For $J=0.8$ of the phase-II, the ground state has a smaller total $S$,
thus $E_g^{S_{max}}-E_g^0>0$ while $E_g^1-E_g^0=0$ representing the same total $S$ states with different $S_z$.
For $J>1.0$, both $E_g^{S_{max}}-E_g^0>0$ and $E_g^1-E_g^0>0$ and the ground state of the whole system is inside the $S=0$ sector.

\textit{Spontaneous chiral current---.} In the ferromagnetic phase, the ground state possesses full spin-polarization.
This system thus reduces to a $t$-$V$ model of spinless fermions. For this model, the system is a semi-metal with quadratic band crossing at the Fermi point in the non-interacting limit of $V=0$. In the presence of a finite interaction $V$, a spontaneous Chern insulator can be established as demonstrated in previous work~\cite{QAHE_Kagome_18}, where the ground states show two-fold degeneracy with opposite chiralities. Here, we demonstrate the spontaneous chiral currents for all three phases. We consider a chiral-symmetry-breaking hopping term $h_{\rm C} = i \chi t_{\rm C} \sum_{\langle ij \rangle, \alpha}c_{i,\alpha}^\dagger c_{j,\alpha}$ as a perturbation, where $t_{\rm C} \ll t=1$ is small and $\chi =\pm1$ when the electron hopping is clockwise/anti-clockwise in each triangle. We can then detect the loop current following the Hellmann-Feynman theorem~\cite{FH_law1, FH_law2, QAHE_QBC_Zeng18} via
\begin{eqnarray}
\langle j \rangle = \frac{1}{2N_s}\frac{\partial E(t_{\rm C})}{\partial {t_{\rm C}}}|_{t_{\rm C} \rightarrow 0},
\end{eqnarray}
where $E(t_{\rm C})=\langle \Psi|H(h_{\rm C})| \Psi \rangle $ with $| \Psi \rangle $ being the ground state of the system and $N_s$ is the number of sites.
In Fig.~\ref{Current}(a), we show the ground state energy difference $E_g^{0}(t_{\rm C})-E_g^{0}(t_{\rm C}=0)$ as a function of $t_{\rm C}$ for $N_s=36$ with $J=0,4$, $0.8$, and $1.4$ representing three different phases, respectively.
One can find that in the ferromagnetic region, the dependence of ground state energy decreases linearly as a function of $t_{\rm C}$ (we use small $t_{\rm C}$ up to $0.008$) corresponding to a loop current of $j=0.028$ for each triangle.
The constant current at weak $J$ is robust against the system size $N_s$, as shown in Fig.~\ref{Current}(b) where the scaling behavior of the current as a function of $1/N_s$ is plotted by the red line with solid circles. We find that the current is finite in the order of $10^{-2}$ at the large $N_s$ limit, suggesting the existence of finite current in the thermodynamic limit. The finite current supports the quantum anomalous Hall effect for phase-I with nonzero $V$.

For partially polarized and nonmagnetic phases with $J=0.8$ and $J=1.4$, the ground state energy decreases near quadratically with $t_{\rm C}$
as shown in Fig.~\ref{Current}(a). For $J=0.8$, the current keeps a small value whereas the current drops a lot at $J=1.4$ as $N_s$ increases.
We caution that because of the quadratic behavior of currents for phase region II and III, we will see much reduced or vanishing current if we take the small $t_{\rm C}$ limit, which is negligible as it is related to an energy difference at the same order as the relative error in DMRG.

\begin{figure}
	\includegraphics[width=8.5 cm]{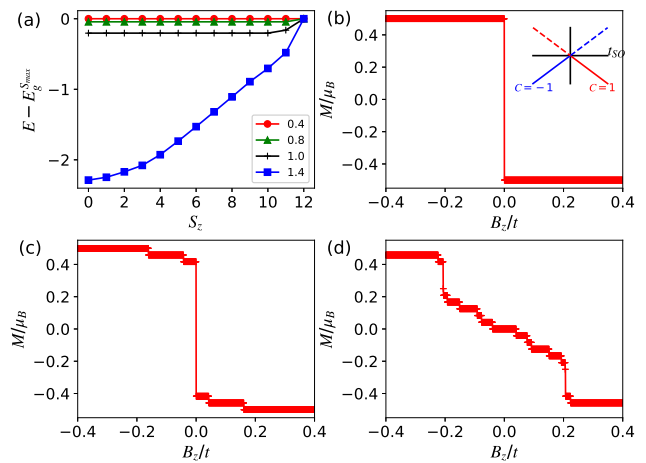}
	\caption{(a) Energy difference between ground state energy at each spin sector and that at maximal spin sector for different $J=0.4$, $0.8$, $1.0$, and $1.4$ for $N_s=72$ site system. (b)-(d) Average magnetization of each electron of the ground state as function of Zeeman field $B_z$ for $J=0.4$, $1.0$, and $1.4$. Inset in (b): Schematic plot of splitting of degenerate ground states with spontaneous QAHE of Chern number $C=\pm1$ due to the presence of SOC as function of its magnitude $t_{\rm SO}$.}
	\label{ZeemanField}
\end{figure}

\textit{Effect of Zeeman field and SOC---.} These quantum phases can be tuned by external Zeeman field. 
In the absence of SOC, the system exhibits SU(2) spin-rotation symmetry. We set the Zeeman field along $z$-direction with $h_Z=B_z \sum_{i}S_i^z$ without loss of generality. 
As $S_z$ is a good quantum number, we can calculate the ground states in each $S_z$ sector as shown in Fig.~\ref{ZeemanField}(a) for different $J$ at $B_z=0$.
One finds that at $J=0.4$, all the ground state energies at different $S_z$ are degenerate and thus they show the same total spin $S=N_e/2$.
The degeneracy indicates that the ground state will be fully polarized along the field direction in the presence of infinitesimally weak perturbation as illustrated in Fig.~\ref{ZeemanField}(b) where the average spin polarization of each unit cell $M$ is plotted as a function of $B_z$.
As $J$ increases to around $0.8 \sim 1.0$, the ground states become partially polarized with total spin $S<N_e/2$. When $J$ is further increased to $1.4$, the degeneracy is completely lifted and the ground state lives in the $S_z=0$ sector as a nonmagnetic phase. For these partially polarized and nonmagnetic phases, the presence of a nonzero Zeeman field can enhance
the magnetization as shown in Figs.~\ref{ZeemanField}(c) and (d), respectively.
In the large Zeeman field limit, the ground state will become fully spin polarized, which also has a chiral current
and quantum anomalous Hall effect since the nature of this state is identical to the spontaneous polarized state (phase I).

In the presence of SOC, the direction of Zeeman field matters. 
The SOC considered here is Kane-Mele type~\cite{SM, 18_Nature_Ye_Dirac, 18_Ye_dHvA_Fe3Sn2, 18_Nature_Hasan_Fe3Sn2, 19_Kang_FeSn_DiracFlat, 19_NatPhys_FlatBandMag_Co3Sn2S2} since the lattice exhibits out-of-plane mirror symmetry. 
When the Zeeman field points out-of-plane, either the SOC~\cite{SM} or the extended Hubbard interaction can induce Chern insulators. The ground state of the former one is nondegenerate while that of the latter one is two-fold degenerate. In the presence of both terms, the two-fold degeneracy between right- and left-handed ground states with Chern numbers of $\pm1$ is lifted as illustrated in the inset of Fig.~\ref{ZeemanField}(b). As a result, the ground states undergoes a topological phase transition with a sign change of Chern number by changing the sign of either SOC or Zeeman field, which is in agreement with the experimentally observed spin-dependent orbital-magnetization in Co$_3$Sn$_2$S$_2$~\cite{19_NatPhys_FlatBandMag_Co3Sn2S2}.

When the magnetization lies in plane along the Zeeman field, we find that the SOC cannot induce Chern insulator phase~\cite{SM} constrained by the joint symmetry of time-reversal operation and out-of-plane mirror reflection~\cite{SM, QAHE_InPlane}. Therefore, only the interaction driven Chern insulator phase can exist with in-plane magnetization. Such difference can be employed to distinguish the physical origin of topological responses in kagome metals.

\textit{Nematic charge density wave---.} To characterize the quantum phase at phase-III, we measure the density, bond, and current of the ground state. We find that the ground state is a non-magnetic insulator with a nematic charge density wave as illustrated below. In Fig.~\ref{DensityCDW}(a), we plot the expectation value of electron number operator $n_i$ at each site, where the circle size is proportional to the electron number $\langle n_i \rangle$. One can find that the charge densities for different sublattices are imbalanced, where A and B sites show similar densities, which are much larger than that of sublattice C. The intra-unit cell charge density difference $\delta n_i=n_{A,i}-n_{C,i}$ at $i$th unit cell is plotted as a function of the unit cell position $i_x$ in the inset of Fig.~\ref{DensityCDW}(b). The charge imbalance away from the boundaries shows weak spatial dependence. Such a charge-density pattern preserves the translation symmetry of the system while breaks the rotation symmetry leading to the nematicity. We have checked that such a density pattern is robust as we increase the width $L_y$ of cylinder systems. The insulating nature of this phase is characterized by a charge excitation gap $\Delta_{2e}=E_g^{0}(N_e+2)+E_g^{0}(N_e-2)-2E_g^{0}(N_e)$ at ground state spin sector of $S_z=0$ by adding/removing two electrons (one spin up and one spin down). The scaling behavior of the charge gap as a function of $1/N_s$ is plotted in Fig.~\ref{DensityCDW}(b), where one can find that the charge gap is finite in thermodynamic limit suggesting the system is an insulator.

Since $J$ represents the strength of the antiferromagnetic coupling, we also check if there is any magnetic order. We show that the state is non-magnetic due to the small site filling number where the antiferromagnetic coupling becomes less efficient. The spin-spin correlation $|\langle \bm{S}_i\cdot {\bm S}_j\rangle|$ decreases exponentially as a function of the distance between two sites as shown in the lower panel of Fig.~\ref{DensityCDW}(c), where the correlation between sites $i$ and $j$ with the same $y$ coordinate is plotted as a function of their distance $i_x-j_x$ along $x$-direction.
We also study the single-particle Green function $G_{ij}\equiv i \langle c_i^\dagger c_j \rangle$ as shown in the upper panel of Fig.~\ref{DensityCDW}(c) where the magnitude of $G_{ij}$ also decreases exponentially as the distance increases, being consistent with a charge insulator state.

\begin{figure}
	\includegraphics[width=8.5 cm]{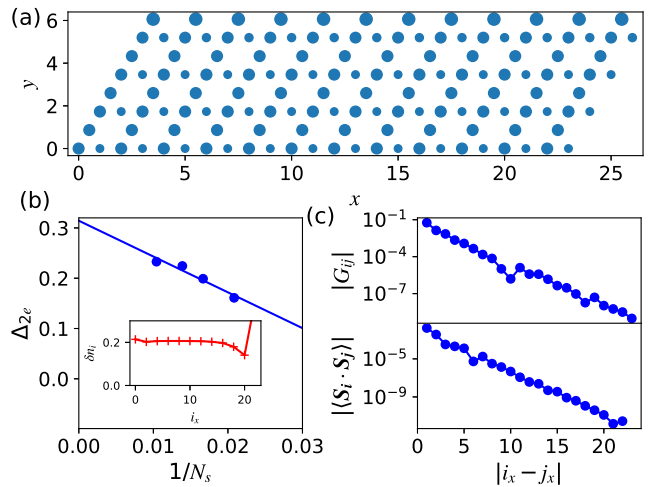}
	\caption{Measurements of ground states in phase region III with $J=1.5$ and $V=0.5$. (a) Charge density at each site. Circle size represents the magnitude of density. (b) Scaling behavior of the charge gap. Inset: intra-unit cell charge density imbalance $\delta n_i$ as the unit-cell position along $x$ direction $i_x$. (c) Dependence of single-particle Green function $G_{ij}$ and spin-spin correlation between two sites on their distance.}
	\label{DensityCDW}
\end{figure}

\textit{Summary and discussion---.} We demonstrated the quantum phase diagram of a metallic kagome lattice with half-filled lowest flat-band by considering the interplay between extended Hubbard repulsion interaction $V$ and the nearest neighbor antiferromagnetic Heisenberg interaction $J$. In the weak $J$ regime, the system exhibits a fully spin-polarized ferromagnetic phase, which hosts a topologically nontrivial spontaneous Chern insulator state in the presence of a finite $V$. 
As $J$ increases gradually, the ferromagnetic topological phase is suppressed and quantum phase transition occurs. The system first becomes a partially spin-polarized phase and then enter a non-magnetic insulating phase with nematic charge density wave order. 

In the fully spin-polarized ferromagnetic phase, the presence of an infinitesimal Zeeman field selects the ground state with spins polarized parallel to the field direction. The spontaneous Chern insulator phase driven by interaction emerges independent of the spin orientation. 
In contrast, the Chern insulator induced by SOC shows a strong dependence on the spin polarization, which disappears when the polarization lies in-plane~\cite{SM}.
Our work provides a way to distinguish the physical mechanisms of the Chern insulator for realistic materials.

One experimental material candidate is FeSn, which shows mirror reflection symmetry about the kagome plane and exhibits flat band close to the Fermi energy~\cite{19_Kang_FeSn_DiracFlat}. 
We suggest that one can measure the anomalous Hall effect or orbital magnetization when the ferromagnetism is aligned in the kagome plane to detect if there is a spontaneous Chern insulator phase driven by interaction. 
Besides, the flat bands have also been observed in kagome metals contributed from other $d$-orbitals~\cite{19_Kang_FeSn_DiracFlat, 19_Kang_FeSn_ElecMagneticProp, 20_CoSn_FlatBand_NonMag} and other atomic layers~\cite{KagomeMetal_Organic_Liu_17, KagomeSiliceneMorie_18, KagomeGraphene_18, KagomeGraphene_19, BoronKagome_19, KagomeOrganic_19}, which provide ideal platforms to test our theory.

\textit{Acknowledgments.---} Y.R. and Z.Q. acknowledge the financial support from the National Key R \& D Program (No. 2017YFB0405703), NNSFC (No. 11974327), Fundamental Research Funds for the Central Universities, and Anhui Initiative in Quantum Information Technologies.
This material is based upon work supported by the U.S. Department of Energy, Office of Science, Advanced Scientific Computing Research and Basic Energy Sciences, Materials Sciences and Engineering Division, Scientific Discovery through Advanced Computing (SciDAC) program under the grant number DE-AC02-76SF00515 (HCJ,DNS).

\end{document}